\documentstyle[prl,aps,preprint,eqsecnum]{revtex}
\tighten 
 
\begin{document}
\draft

\def\beq{\begin{equation}}
\def\eeq{\end{equation}}
\def\eeql#1{\label{#1} \end{equation}}
\def\bea{\begin{eqnarray}}
\def\eea{\end{eqnarray}}
\def\eeal#1{\label{#1} \end{eqnarray}}
\def\phs{{\vphantom{*}}}
\def\hn{\mskip-0.5\thinmuskip}
\def\hp{\mskip0.5\thinmuskip}
\renewcommand{\vec}{\bbox}
\def\im{\mathop{\rm Im}\nolimits}
\def\re{\mathop{\rm Re}\nolimits}
\def\fj{\vec{f}_{\!j}}
\def\ds{\displaystyle}

\title{Exactly solvable path integral for open cavities\\ in terms of quasinormal modes}

\author{Alec Maassen van den Brink}

\address{Department of Physics, The Chinese University of Hong Kong,
Hong Kong, China}

\date{\today}
\maketitle

\begin{abstract}

We evaluate the finite-temperature Euclidean phase-space path integral for the generating functional of a scalar field inside a leaky cavity. Provided the source is confined to the cavity, one can first of all integrate out the fields on the outside to obtain an effective action for the cavity alone. Subsequently, one uses an expansion of the cavity field in terms of its {\em quasinormal modes\/} (QNMs)---the exact, exponentially damped eigenstates of the classical evolution operator, which previously have been shown to be complete for a large class of models. Dissipation causes the effective cavity action to be nondiagonal in the QNM basis. The inversion of this action matrix inherent in the Gaussian path integral to obtain the generating functional is therefore nontrivial, but can be accomplished by invoking a novel QNM sum rule. The results are consistent with those obtained previously using canonical quantization.

\end{abstract}

 


\section{Introduction}

Open systems have been the subject of numerous investigations both in classical and quantum physics. Physical examples occur for instance in optics---e.g., in cavity QED\cite{optcav} with obvious applications to laser physics\cite{meystre}, and in microdroplets, which provides the spherically symmetric analog---and solid-state physics, where Josephson phenomena\cite{S&Z} and Kondo and related problems\cite{weiss-boek} are all amenable to a ``system--bath" description. The open-system concept is also relevant to elastic waves and acoustics (e.g., sound emanating from musical instruments) and on a very different scale to astrophysics and gravitational waves\cite{grav}.

In a series of papers, we have been particularly interested in those open systems which are governed by a wave equation (Eq.~(\ref{wave-eq}) below or, by a simple transformation\cite{comp2}, the Klein--Gordon equation $[\partial_t^2-\partial_x^2+V(x)]\psi(x,t)=0$). The ``universe" in which the waves propagate is supposed to consist of a ``cavity" (the open system) and an infinite ``outside", and dissipation is caused by waves escaping from the former to the latter\cite{HD-rep}. Attention has been focused on the role of the cavity resonances or {\em quasinormal modes\/} (QNMs)---eigensolutions of the (non-Hermitian) evolution operator of the open system which are exponentially decaying in time. It turns out that under certain conditions, to be specified later, the discrete set of QNM wavefunctions is complete in the cavity region of space and hence can be used for exact eigenfunction expansions. This eliminates the outside from the description; in particular, one no longer has to deal with the dense set of modes of the infinite universe. In terms of these QNMs, one can establish a formalism which closely parallels the usual one for eigenfunction expansions in conservative, Hermitian systems. Applications of this formalism include perturbation theory and (canonical) second quantization\cite{KL,sq}, the latter being of particular interest to us here. These developments are reviewed in Ref.\cite{RMP}.

Of course, there are many other ways to eliminate the outside (or bath) and obtain an effective description in terms of the damped degrees of freedom of the system only, leading to, {\em inter alia}, Langevin and master equations\cite{meystre,plank}. A formulation which has proved to be especially suited for the study of open quantum systems is the path integral\cite{weiss-boek,diss}. First, one writes down a functional integral for the generating functional (or density matrix) of the universe; the pertinent action, which is readily deduced from the Hamiltonian, is supposedly known. Subsequently, one performs the integral over the bath degrees of freedom only, in which the system (or cavity) coordinates figure as constant parameters. Since the bath is usually taken to be harmonic (in fact, a meaningful separation into system plus bath depends on the latter having this or some other simplifying property), this step can be performed in closed form. One is left with a path integral over the system degrees of freedom, in which the integrand involves an {\em effective action\/} which accounts for the environment in a way that is guaranteed to be consistent with quantum mechanics (as opposed to being merely phenomenological). This reduced path integral then is a convenient starting point for approximations, qualitative considerations or numerical evaluation.

This paper's aim is, in short, to initiate a synthesis between the QNM and path-integral approaches to open systems characterized by escaping waves. To this end, in Section~\ref{class} we first review the basics of the classical QNM expansion for such wave systems to make the paper self-contained, and to establish some notation. In Section~\ref{elim} we present the (finite-temperature Euclidean phase-space) path integral for a simple class of wave systems. Since our ultimate interest is in quantities involving the cavity fields, the source function in the generating functional is chosen to couple to those fields only, which facilitates the subsequent elimination of the bath. This step does not yet involve the QNMs, and the ensuing effective action (in particular, the damping term) turns out to have a form which is still well known.

In Section~\ref{cavQNM}, the completeness of the QNMs for the models considered is used to write this effective action in terms of the QNM basis. This brings together the advantages of an effective action with those of a discrete basis\cite{RMP}. Up to this stage, the calculation has only used the assumption that the outside is harmonic. However, assuming the cavity to be harmonic as well, it turns out that substantially more can be done: the special properties of the QNMs---in particular, a novel sum rule which we also derive---enable the cavity integral to be performed, as an integral over the QNM expansion coefficients. (If the cavity action is nonlinear, this integral over the harmonic part is the starting point for the perturbation expansion; see Section~\ref{discuss}.) Since the effective action is bilinear in the fields and since the QNMs are not orthogonal in the usual sense, this involves inversion of a non-diagonal infinite matrix (this is to be contrasted with the integral over the outside fields, which can be performed for each degree of freedom separately). In our non-interacting problem, this is the nontrivial step, brought about by the nontrivial mass density distribution $\rho(x)$ in (\ref{wave-eq}) below. The result is found to be in agreement with the result of canonical quantization, in that both methods yield the same correlation functions. While the damped systems in Refs.\cite{S&Z,weiss-boek,diss} typically have {\em few\/} degrees of freedom, only their baths being essentially infinite, this paper carries out the analogous program of path-integral quantization for a damped {\em field}.

In parallel with a development in the canonical quantization of the system, in Section~\ref{one-comp} we discuss the special case that the external source couples to the scalar field but not to its conjugate momentum. Then, the aforementioned infinite matrices {\em can\/} be transformed into a diagonal form, i.e., one involving only a single sum over QNM indices. While the resulting expressions look simpler, limitations to their applicability are pointed out. Finally, some closing remarks are made in Section~\ref{discuss}.

The appendices contain supplementary material. Since the subjects of Appendices \ref{sumApp} and~\ref{diagApp} are not specific to path integration, they are of independent interest for the study of QNMs.

\section{Classical fields}
\label{class}

In this section, we summarize the QNM expansion for classical fields. In this paper, we deal with scalar fields in 1 d only.

For closed, linear systems, eigenfunction expansions, based on the eigenfunctions or normal modes (NMs) of their evolution operators, are a tool of vital importance in theoretical physics.  However, open systems are not directly amenable to an NM analysis. In these systems, any initial state decays in time, so stationary NMs do not exist.  As the simplest example, we shall be concerned with the real scalar wave equation in one space dimension,
\beq
  \rho(x)\hp\partial_t^2\phi = \partial_x^2\phi
\eeql{wave-eq}
studied in a cavity $0\le x\le a$, with the nodal boundary condition
\beq
  \phi(x{=}0,t)= 0
\eeql{bc-left}
at one end but with the outgoing-wave condition (OWC)
\beq
  \phi'(a^+,t) = -\dot{\phi}(a,t)
\eeql{bc-right}
at the other. The OWC states that, just outside the cavity boundary, the field $\phi(x,t)$ is an outgoing wave $\phi(x-t)$ (cf.\ (\ref{rho-out}) below); the condition is specified at $a^+$ because, as we shall see below, one is often concerned with models in which there is a singularity in $\rho(x)$ at $x=a$, leading to a possible discontinuity in $\phi'(x)$ \cite{phi-cont}.  The boundary condition (\ref{bc-right}) turns the cavity into a dissipative system that is leaky but not absorptive.  The model (\ref{wave-eq}) has been widely used as the scalar model of electromagnetism in an optical cavity \cite{optcav}.  More physically, the 1-d nature is realized in Fabry--Perot cavities with lengths much smaller than the lateral dimensions, and the scalar-field model is rigorously applicable to the transverse electric sector.

For the open system (\ref{wave-eq})--(\ref{bc-right}), the eigensolutions, labeled by an index $j$, have the form
\beq
  \phi(x,t) = f_j(x)\hp e^{-i\omega_jt}\;,
\eeql{eigensol}
with the QNMs or cavity resonances $f_j$ satisfying
\beq
  [\partial_x^2+\rho(x)\hp\omega_j^2]\hp f_j = 0
\eeql{def-QNM}
and the boundary conditions (\ref{bc-left}), (\ref{bc-right}) translating to
\beq
  f_j(0) = 0,  \qquad   f_j'(a^+) = i\omega_jf_j(a)\;.
\eeql{f-bc}
It is easily verified that $\im\omega_j <0$, so that the solution (\ref{eigensol}) is indeed decaying in time.  Furthermore, the frequencies $\omega_j$, which we suppose to be ordered according to increasing real parts, are spaced by $\Delta\omega\sim\pi/a$, approximately as for a conservative system of size $a$.  With the possible exception of modes with $\re\omega_j =0$, the QNMs always occur in pairs with $\omega_{-j}^\phs=-\omega_j^*$, and one can choose $f_{-j}^\phs=f_j^*$. While the field $\phi$ is real, the eigenvalues and eigenfunctions are complex; this is the reason for the pairing of modes.

The usual formalism concerning eigenfunction expansions relies on the hermiticity of the evolution operator, which only holds in the conservative case, and therefore breaks down for open systems. One possible resolution is to embed the cavity into a universe $0\le x\le\Lambda$ with a nodal condition at $x=\Lambda\rightarrow\infty$, and study its NMs---the modes of the universe. Namely, the system (\ref{wave-eq})--(\ref{bc-right}) is the restriction to $x\le a$ of the problem (\ref{wave-eq}) on the half line $0\le x<\infty$, if one sets
\beq
  \rho(x{>}a) \equiv 1
\eeql{rho-out}
and with the extension of the initial conditions to the ``outside" $x>a$ obeying $\phi'(x{>}a,t{=}0)=-\dot{\phi}(x{>}a,t{=}0)$.  However, this has the obvious disadvantage of having to work with a continuum of states (spaced by $\Delta \omega \sim \pi / \Lambda \rightarrow 0$) as opposed to the discrete set of eigenfunctions in the conservative case.  Besides, the closed system of equations (\ref{wave-eq})--(\ref{bc-right}) shows that even in the presence of dissipation the (thermo-)dynamics of the cavity can be studied {\em without}\/ explicit reference to the outside, which is the principal goal of the program of second quantization of the open system.

Previous work (see \cite{RMP} and references therein) has established that, in spite of the lack of Hermiticity in the conventional sense, an eigenfunction expansion for outgoing waves in classical open wave systems can be formulated in terms of the cavity degrees of freedom only, overcoming the disadvantages of the modes of the universe approach. The sufficient conditions for this QNM expansion are as follows.

\begin{itemize}
\item[(a)]The function $\rho(x)$ has at least a kink discontinuity at $x=a$.  This demarcates a well-defined cavity region from the rest of the universe. In fact, at several places the type of discontinuity will be of relevance to us, and we define a parameter $\mu$ by $\rho(x)=\rho_{\rm reg}(x)+\mu\hp\delta(x-a)$, where $\rho_{\rm reg}(x)$ has at most a step discontinuity at $x=a$.
\item[(b)]The function $\rho(x)$ has no tail outside the cavity, i.e., $\rho(x{>}a)\equiv1$. This condition ensures that the outside does not reflect outgoing waves back into the cavity, enabling the complete elimination of the environment from the equations of motion.
\end{itemize}
These conditions are satisfied for optical cavities bounded from extended vacuum by a sharp material interface. Under these conditions, the eigenfunction expansion is exact for any amount of dissipation.

The completeness of the QNMs can be pursued at two levels. First, one shows that the retarded Green's function of the system has the representation\cite{comp1}
\begin{equation}
  G^{\rm R}(x,y;t) =\sum_j \frac{f_j(x)f_j(y)}{2i\omega_j}\hp e^{-i\omega_jt}
\eeql{greenfie}
for $0 \le x,y \le a$ and $t \ge 0$, where the $f_j$'s are normalized according to (\ref{normal}) below. Thus, the dynamics is contained entirely in the QNMs.

Second, realizing that the wave equation (\ref{wave-eq}), like any classical Hamiltonian problem, requires both position and momentum to be specified as initial data, one introduces function {\em pairs\/} $\vec{\phi}=(\phi,\hat{\phi})^{\rm T}$ with the conjugate momentum $\hat{\phi}\equiv\rho\dot{\phi}$, so that for eigenfunctions $\fj=(f_j,-i\rho\omega_jf_j)^{\rm T}$\cite{tong}. The set of all function pairs (in general allowed to be complex) satisfying the boundary conditions (\ref{bc-left}) and~(\ref{bc-right}) will be denoted as $\Gamma$---the space of outgoing waves. A report on the formal mathematical construction is currently in preparation\cite{factor}.

Using these pairs, one can prove that the time evolution generated by (\ref{greenfie}) can be recast in the form
\beq
  \vec{\phi}(t) = \sum_j a_j(t) \fj\;,
\eeql{evolve}
where the expansion coefficients are given by
\beq
 a_j(t) = \frac{1}{2\omega_j}(\fj,\vec{\phi}(t))
\eeql{a_n1}
with $a_j(t)=a_j(0)\hp e^{-i\omega_jt}$ and the {\em bilinear map\/} for $\vec{\zeta},\vec{\chi}\in\Gamma$
\beq
  (\vec{\zeta},\vec{\chi}) = i \biggl\{
  \int_0^{a^+}\!\!dx\,\left[ \zeta(x)\hp\hat{\chi}(x)
  +\hat{\zeta}(x)\hp\chi(x)\right]
  +\zeta(a)\hp\chi(a) \biggr\} \;.
\eeql{scalarprod}
By simply letting $t\!\downarrow\!0$ in (\ref{evolve}) one arrives at a {\em two-component expansion\/} for an arbitrary real $\vec{\phi}\in\Gamma$.  This expansion makes the completeness of the QNMs manifest.  The normalization used in (\ref{greenfie}) to (\ref{a_n1}) can be concisely expressed as
\beq
  (\fj,\fj)=2\omega_j\;.
\eeql{normal}
It is seen that (\ref{normal}) in general is not real, underlining the difference between the form (\ref{scalarprod}) and a conventional scalar product involving complex conjugation.  The fact that (\ref{normal}) is bilinear also serves to establish a phase convention for the wavefunctions.

Upon introducing the two-component evolution operator
\beq
  {\cal H}=i \pmatrix{ 0 & \rho(x)^{-1} \cr \partial_x^2 & 0}\;,
\eeql{harray}
the cavity evolution (\ref{wave-eq}) can be written as $i\hp\partial_t\vec{\phi}={\cal H}\vec{\phi}$, in striking analogy with quantum mechanics. In this notation, the definition (\ref{def-QNM}) of $f_j$ takes the form ${\cal H}\fj=\omega_j\fj$. The operator ${\cal H}$ can be shown to be symmetric with respect to the form (\ref{scalarprod}), i.e.,
\beq
  (\vec{\zeta},{\cal H}\vec{\chi}) =
  ({\cal H}\vec{\zeta},\vec{\chi})
\eeq
for any $\vec{\zeta},\vec{\chi} \in \Gamma$. This analog of hermiticity holds even though the system is not conservative. The symmetry of ${\cal H}$ yields the ``orthogonality" relation
\beq
  (\fj,\vec{f}_{\hn\!k})=0 \qquad\text{for $\omega_j\neq\omega_k$}
\eeql{ortho}
in an immediate transcription of the usual proof, leading to the uniqueness of the expansion.  Incidentally, an expansion such as (\ref{evolve}) but involving the first component alone would not be unique.

Instead of its present formulation as an ``orthogonal" expansion involving a bilinear map, the series (\ref{evolve}) can also be regarded as a bi-orthogonal expansion in terms of the standard inner product\cite{bior}. The power of this latter, slightly more involved method shows when several QNMs merge\cite{jordan}, a possibility which in this paper we will only briefly consider in Appendix~\ref{JBpath}.

\section{Elimination of the outside}
\label{elim}

We want to calculate the generating functional for the cavity field, in a form which manifestly involves the QNMs. Since we want results for finite temperature, a Euclidean formulation is advantageous. The QNM expansion (\ref{evolve}) involves two components; therefore, we have to use the phase-space version of the path integral:
\bea
  S\{\vec{\chi}\}&=&\left\langle{\cal T}_\tau\exp\biggl\{\int_0^\beta\!\!d\tau\,
  \bigl(\vec{\phi}(\tau),\vec{\chi}(\tau)\bigr)\biggr\}\right\rangle
    \label{expect}\\
  &=&Z^{-1}\!\int\!{\cal D}\vec{\phi}(x,\tau)\exp\left\{
  \int_0^\beta\!\!d\tau\left[(\vec{\phi},\vec{\chi})
  -\int_0^{a+\Lambda}\!\!dx\left(\frac{\hat{\phi}^2}{2\rho}
  +\frac{{\phi'}^2}{2}-i\hat{\phi}\dot{\phi}\right)\right]\right\}.
\eeal{pathint}
In (\ref{expect}), $\vec{\chi}$ is a real external source\cite{Imsource}; the coupling to the cavity field only (cf.\ the Introduction) is taken to be of the form (\ref{scalarprod}), anticipating that, by (\ref{ortho}), this will simplify expressions upon QNM expansion as in (\ref{evolve}). The inverse temperature is $\beta=1/T$ (throughout $\hbar=k_{\rm B}=1$). Imaginary-time ordering is denoted by ${\cal T}_\tau$ as usual; this operation is needed in (\ref{expect}), in which $\vec{\phi}$ is still a Heisenberg operator field, but has no place in (\ref{pathint}) which involves $c$-numbers only. In the following, the meaning of $\vec{\phi}$ will be clear from the context. The two-component measure is ${\cal D}\vec{\phi}\equiv{\cal D}\phi\hp{\cal D}\hat{\phi}$. The normalization $Z$ equals the path integral on the r.h.s.\ with $\vec{\chi}\mapsto0$. This is to be understood formally, for in fact both the integral and $Z$ are infinite, only their ratio being meaningful. In the course of the calculation we shall cancel all factors which do not depend on $\vec{\chi}$ against corresponding factors in $Z$, without bothering to reflect these subsequent redefinitions of $Z$ in the notation.

The boundary conditions on the real field $\vec{\phi}$ are $\phi(0,\tau)=\phi(a+\Lambda,\tau)=0$, $\phi(x,0)=\phi(x,\beta)$, the latter arising from the trace implicit in the quantum statistical expectation value (\ref{expect}). No boundary conditions can be imposed on $\hat{\phi}$, since the absence of terms with $\hat{\phi}'$ in the action implies that $\hat{\phi}$ typically is completely discontinuous. This at the same time means that phase-space path integrals can be quite tricky\cite{schulman,roepstorff}. However, these problems in general show up only beyond the semi-classical approximation, and hence should be absent here since we look at a linear problem for which this approximation is exact\cite{inter}. We are interested in the limit $\Lambda\rightarrow\infty$, but postpone this operation until it can be performed in a controlled way.

We want to split the integral into a cavity and a bath factor: $Z^{-1}\!\int\!{\cal D}\vec{\phi}=Z_{\rm c}^{-1}\!\int\!{\cal D}\vec{\phi}_{\rm c}\*\,Z_{\rm b}^{-1}\!\int\!{\cal D}\vec{\phi}_{\rm b}$, where the latter runs over fields on $(a,a+\Lambda)$, with a given boundary value $\phi(a,\tau)$. The integral over bath momenta is trivial; introducing $\xi\equiv x-a$,
\beq
  \int\frac{{\cal D}\hat{\phi}_{\rm b}}{Z_{\rm b}}\exp\left\{
  -\int_0^\beta\!\!d\tau\!
  \int_0^{\Lambda}\!\!d\xi\left(\frac{\hat{\phi}^2}{2}
  +\frac{{\phi'}^2}{2}-i\hat{\phi}\dot{\phi}\right)\right\}=
  \exp\left\{-\frac{1}{2}\int_0^\beta\!\!d\tau\!
  \int_0^{\Lambda}\!\!d\xi\left(\dot{\phi}^2+{\phi'}^2\right)\right\},
\eeql{phibhat-int}
where we used (\ref{rho-out}). The $\phi_{\rm b}$-integral is nontrivial only because of the boundary condition, which can be implemented through the expansion
\beq
  \phi_{\rm b}(\xi,\tau)=T\sum_m\left\{\phi_m(a)\frac{\Lambda-\xi}{\Lambda}
  +\sum_{u=1}^{\infty}\phi_{um}\sin\left(\frac{\pi u\xi}{\Lambda}\right)
  \right\}e^{-i\nu_m\tau},
\eeql{phi-b}
with the bosonic Matsubara frequencies $\nu_m=2\pi mT$, $m\in{\bf Z}$.

Substituting (\ref{phi-b}) into (\ref{phibhat-int}), the action can be written as
\bea
  \frac{1}{2}&&\int_0^\beta\!\!d\tau\!
  \int_0^{\Lambda}\!\!d\xi\left(\dot{\phi}^2+{\phi'}^2\right)
  \nonumber\\&&=
  \frac{T}{2}\sum_m\left\{\!\left(\frac{\Lambda\nu_m^2}{3}
  +\Lambda^{-1}\right)\!|\phi_m(a)|^2+\sum_{u=1}^{\infty}\left[\!\left(
  \frac{\Lambda\nu_m^2}{2}+\frac{\pi^2u^2}{2\Lambda}\right)\!|\phi_{um}|^2
  +\frac{2\Lambda\nu_m^2}{\pi u}\re(\phi_{um}\phi_{-m}(a))\!\right]\!
  \right\}\nonumber\\
  &&=\frac{T}{2}\sum_m\left\{\!\left(\frac{\Lambda\nu_m^2}{3}
  +\Lambda^{-1}\right)\!|\phi_m(a)|^2
  +\sum_{u=1}^{\infty}\left[\!\left(
  \frac{\Lambda\nu_m^2}{2}+\frac{\pi^2u^2}{2\Lambda}\right)
  \!|\bar{\phi}_{um}|^2
  -\frac{2(\Lambda\nu_m^2/\pi u)^2|\phi_m(a)|^2}{\Lambda\nu_m^2
  +\pi^2u^2/\Lambda}\right]\!\right\},\nonumber\\
  &&
\eeal{complsq-phi}
where
\beq
  \bar{\phi}_{um}=\phi_{um}+
  \frac{2\nu_m^2\phi_m(a)/\pi u}{\nu_m^2+(\pi u/\Lambda)^2}\;.
\eeq
Switching to variables $\bar{\phi}_{um}$ does not affect the domain of integration (in particular not in a $\phi_m(a)$-dependent way), since both the $\phi_{um}$ and the $\bar{\phi}_{um}$ run over all $\bf C$, subject only to the restriction $\phi_{um}^\phs=\phi_{u,-m}^*$ and analogously for $\bar{\phi}_{um}$. The completion of the square in the last line of (\ref{complsq-phi}) thus eliminates the need to actually perform the path integral, and one obtains some constant independent of $\phi(a)$, which then cancels against the same factor in $Z_{\rm b}$. This leads to
\bea
  \int\frac{{\cal D}\phi_{\rm b}}{Z_{\rm b}}&&\,\exp\left\{
  -\frac{1}{2}\int_0^\beta\!\!d\tau\!
  \int_0^{\Lambda}\!\!d\xi\left(\dot{\phi}^2+{\phi'}^2\right)\right\}\nonumber\\
  &&=\exp\left\{\frac{T}{2}\sum_m|\phi_m(a)|^2\left[-\frac{\Lambda\nu_m^2}{3}
  -\Lambda^{-1}+\sum_{u=1}^{\infty}\frac{2(\Lambda\nu_m^2/\pi u)^2}
  {\Lambda\nu_m^2+\pi^2u^2/\Lambda}\right]\right\}\label{sum-n}\\
  &&=\exp\left\{-\frac{T}{2}\sum_m|\nu_m||\phi_m(a)|^2\right\}\quad
  \mbox{for $\Lambda\rightarrow\infty$},
\eeal{CL-damp}
where to arrive at the last line we evaluated the sum over $u$ asymptotically in $\Lambda^{-1}$, using
\bea
  \sum_{u=1}^{\infty}\frac{1}{u^2(1+\epsilon^2u^2)}&=&
  \sum_{u=1}^{\infty}\left(\frac{1}{u^2}-\frac{1}{u^2+\epsilon^{-2}}\right)
  \nonumber\\&=&\frac{\pi^2}{6}-\int_0^{\infty}\!\frac{du}{u^2+\epsilon^{-2}}
    +{\cal O}(\epsilon^2)\nonumber\\
  &=&\frac{\pi^2}{6}-\frac{\pi|\epsilon|}{2}+{\cal O}(\epsilon^2)
\eeal{asympt}
for $\epsilon=\pi/\Lambda\nu_m$, leading to the cancellation of the ${\cal O}(\Lambda)$ terms in the exponent of (\ref{sum-n}).

The Caldeira--Leggett type\cite{weiss-boek} exponent in (\ref{CL-damp}) is the quantum-mechanical, finite-temperature equivalent of an Ohmic-damping term. Its emergence here is not surprising, given the correspondence between transmission-line environments of the type considered here\cite{HDstring} and the oscillator baths used in its original derivation\cite{diss}. Namely, in the limit $\Lambda\rightarrow\infty$ waves escaping into the homogeneous outside string will never be scattered back, so that the outside acts as a sink for waves emanating from the cavity. Since the model (\ref{wave-eq}) is dispersionless, this damping is frequency independent. On the classical level this leads to (\ref{bc-right}), in which on the l.h.s.\ $\phi'$ is precisely the string tension; according to the r.h.s., this force is equal to $-\dot{\phi}$. This velocity-proportionality is reflected by the first power of $\nu_m$ in (\ref{CL-damp}); however, unlike (\ref{bc-right}), the action (\ref{CL-damp}) is necessarily (imaginary-)time reversal invariant.

Substituting the bath contribution back into the generating functional (\ref{pathint}) and transforming to Bose frequencies also in the cavity, one gets
\bea
  S\{\vec{\chi}\}=\int\frac{{\cal D}\vec{\phi}_{\rm c}}{Z}\,\exp\Biggl\{
  T\sum_m\Biggl[&&(\vec{\phi}_m,\vec{\chi}_{-m})-\frac{|\nu_m|}{2}|\phi_m(a)|^2
  \nonumber\\&&
  -\int_0^a\!\!dx\left(\frac{|\hat{\phi}_m|^2}{2\rho}+\frac{|\phi_m'|^2}{2}
  -\nu_m\phi_m\hat{\phi}_{-m}\right)\Biggr]\Biggr\}\;.
\eeal{eff-act}
This form completes the elimination procedure in that it manifestly involves $\vec{\phi}_{\rm c}$ only.

\section{Performing the cavity-field integral in the QNM basis}
\label{cavQNM}

\subsection{Path integral and Green's function}
\label{cav-int}

To make further progress, we introduce the QNM expansions $\vec{\phi}_m=\sum_ja_{jm}\fj$ and $\vec{\chi}_m=\sum_jb_{jm}\fj$\cite{out-exp}; the coefficients satisfy $a_{jm}^\phs=a_{-j,-m}^*$, $b_{jm}^\phs=b_{-j,-m}^*$. Substitution into (\ref{eff-act}) and invoking the ``orthogonality" relation
\beq
  \int_0^{a^+}\!\!dx\,\rho f_jf_k=
  \delta_{jk}-i\hp\frac{f_j(a)f_k(a)}{\omega_j+\omega_k}\;,
\eeql{ortho2}
which follows from (\ref{scalarprod}) and (\ref{ortho}), leads to
\bea
  S\{\vec{\chi}\}&=&\int\frac{{\cal D}\vec{\phi}_{\rm c}}{Z}\,\exp\Biggl\{
  T\sum_m\Biggl[-\sum_{jk}a_{jm}a_{k,-m}f_j(a)f_k(a)
  \frac{\nu_m\left(\theta(m)\hp\omega_k-\theta(-m)\hp\omega_j\right)
  +i\omega_j\omega_k}{\omega_j+\omega_k}\nonumber\\
  &&\hphantom{\int\frac{{\cal D}\vec{\phi}_{\rm c}}{Z}\,\exp\Biggl\{
  T\sum_m\Biggl[}
  +\sum_j2\omega_jb_{j,-m}a_{jm}\Biggr]\Biggr\}\label{S-QNM}\\
  &=&\int\frac{{\cal D}\vec{\phi}_{\rm c}}{Z}\,\exp\Biggl\{
  T\sum_m\Biggl[-\sum_{jk}\bar{a}_{jm}\bar{a}_{k,-m}f_j(a)f_k(a)
  \frac{\nu_m\left(\theta(m)\hp\omega_k-\theta(-m)\hp\omega_j\right)
  +i\omega_j\omega_k}{\omega_j+\omega_k}\nonumber\\
  &&\hphantom{\int\frac{{\cal D}\vec{\phi}_{\rm c}}{Z}\,\exp\Biggl\{
  T\sum_m\Biggl[}
  -\sum_{jk}b_{jm}b_{k,-m}\frac{f_j(a)f_k(a)}{\omega_j+\omega_k}
  \left(\frac{\theta(m)\hp\omega_k}{i\omega_k+\nu_m}
  +\frac{\theta(-m)\hp\omega_j}{i\omega_j-\nu_m}\right)\Biggr]\Biggr\}\;,
\eeal{complsq-a}
where
\beq
  \bar{a}_{jm}=a_{jm}+\sum_k\omega_kb_{km}
  \frac{f_j(a)f_k(a)}{\omega_j+\omega_k}
  \left(\frac{\theta(m)}{\omega_k(i\omega_j+\nu_m)}
  +\frac{\theta(-m)}{\omega_j(i\omega_k-\nu_m)}\right)
\eeql{abar-a}
and $\theta(0)\equiv\frac{1}{2}$. To arrive at (\ref{S-QNM}), one has to use the fact that the Kronecker term in (\ref{ortho2}) does not contribute in the action by $m$-parity. Thus, it is precisely the second term in (\ref{ortho2}) which contributes; its presence makes (\ref{S-QNM}) non-diagonal (i.e., involving a double sum $\sum_{jk}$); the surface values $f_j(a)f_k(a)$ are a measure of dissipation, since they would vanish if the system satisfied a nodal boundary condition also at $x=a$\cite{conslim}.

The only nontrivial ingredient in the completion of the square (\ref{complsq-a}) is the QNM sum rule
\beq
  \sum_k\frac{f_j(a)f_k^2(a)f_{\ell}(a)}
  {(\omega_j+\omega_k)(\omega_k+\omega_{\ell})}=-\delta_{j\ell}\;,
\eeql{sumrule}
which will be discussed in Section~\ref{sumrule-sect}.

We thus arrive at the final answer for the generating functional in terms of the coefficients $b_{jm}$\cite{domain},
\beq
  S\{\vec{\chi}\}=\exp\Biggl\{
  iT\sum_{jkm}b_{jm}b_{k,-m}\frac{f_j(a)f_k(a)}{\omega_j+\omega_k}
  \left(\theta(m)\frac{\omega_k}{\omega_k-i\nu_m}
  +\theta(-m)\frac{\omega_j}{\omega_j+i\nu_m}\right)\Biggr\}\;.
\eeql{Sres}
The relation between $S\{\vec{\chi}\}$ and the temperature Green's function ${\cal G}_{jk}(\tau)=-\left<{\cal T}_\tau\{a_j(\tau)\hp a_k\}\right>$ is standard\cite{diff-b}:
\bea
  \left.\frac{\partial^2S\{\vec{\chi}\}}
    {\partial b_{jm}\partial b_{k,-m}}\right|_{\vec{\chi}=0}&=&
  \left.\frac{\partial^2}
    {\partial b_{jm}\partial b_{k,-m}}\right|_{\vec{\chi}=0}\left<{\cal T}_\tau
  \exp\biggl\{T\sum_{jm}2\omega_jb_{jm}\int_0^\beta\!\!d\tau\,e^{-i\nu_m\tau}
    a_j(\tau)\biggr\}\right>\nonumber\\
  &=&4\omega_j\omega_kT^2\!\int_0^\beta\!\!d\tau_1\hp d\tau_2\,
     e^{i\nu_m(\tau_2-\tau_1)}
     \left<{\cal T}_\tau\{a_j(\tau_1)\hp a_k(\tau_2)\}\right>\nonumber\\
  &=&-4\omega_j\omega_kT\tilde{\cal G}_{jk}(\nu_{-m})\;,
\eeal{diffS}
where it is a standard result of functional integration\cite{ryder} that differentiation (in our case ordinary partial differentiations with respect to the {\em discrete\/} basis $\{b_{jm}\}$) of the generating functional automatically yields time-ordered expectation values (cf.\ below (\ref{pathint})). Substituting (\ref{Sres}) into (\ref{diffS}), one obtains
\beq
  \tilde{\cal G}_{jk}(\nu_m)=-i\frac{f_j(a)f_k(a)}
  {2\omega_j\omega_k(\omega_j+\omega_k)}\left\{\theta(m)
  \frac{\omega_j}{\omega_j-i\nu_m}
  +\theta(-m)\frac{\omega_k}{\omega_k+i\nu_m}\right\}\;.
\eeql{calG}
Quite generally the temperature Green's function is related to the real-time retarded propagator $G^{\rm R}_{jk}(t)=-i\theta(t)\hp\langle[a_j(t),a_k]\rangle$ by
\beq
  \tilde{\cal G}_{jk}(\nu_m)=\tilde{G}^{\rm R}_{jk}(i\nu_m)
\eeql{GtildeGR}
for $m\ge1$\cite{AGD}. Evaluating the analytically continued $\tilde{G}^{\rm R}_{jk}$ at the discrete frequencies $i\nu_m$, (\ref{calG}) thus is readily seen to be in exact agreement with the results obtained in Ref.\cite{sq} by canonical quantization.

\subsection{Action-inverting sum rule}
\label{sumrule-sect}

In the previous subsection we deferred the derivation of (\ref{sumrule}) not to disrupt the flow of the argument; it will be discussed presently. For $j\neq\ell$ one can apply a partial-fraction expansion to the summand and see that the sum indeed vanishes if $\sum_kf_k^2(a)/(\omega_k+\omega_j)$ is independent of~$j$. This can be shown by rewriting
\bea
  \sum_k\frac{f_k^2(a)}{\omega_k+\omega_j}&=&
  \sum_kf_k^2(a)\left(\frac{1}{\omega_k}
    -\frac{\omega_j}{\omega_k(\omega_k+\omega_j)}\right)\nonumber\\
  &=&\sum_k\frac{f_k^2(a)}{\omega_k}+2\omega_j
     \tilde{G}^{\rm R}(a,a;-\omega_j)\nonumber\\
  &=&\sum_k\frac{f_k^2(a)}{\omega_k}+i\;,
\eeal{offd-sum}
where the last line follows because more generally one has
\beq
  2\omega_j\tilde{G}^{\rm R}(x,a;-\omega_j)=i\frac{f_j(x)}{f_j(a)}\;.
\eeql{GminOj}
For a proof, let $\omega\rightarrow-\omega_j$ in the purely classical identity $\tilde{G}^{\rm R}(x,y;\omega)-\tilde{G}^{\rm R}(x,y;-\omega)=(2\omega/i)\hp\tilde{G}^{\rm R}(x,a;\omega)\hp\tilde{G}^{\rm R}(y,a;-\omega)$ established in Ref.\cite{sq}, and compare residues on both sides. The value of the first term on the r.h.s.\ of (\ref{offd-sum}) is irrelevant for the derivation of (\ref{sumrule}); in Appendix~\ref{sumApp} it will be shown that, for the case of a step discontinuity in $\rho(x{=}a)$, one has $\sum_kf_k^2(a)/(\omega_k+\omega_j)=i(\rho(a^-)+1)/(\rho(a^-)-1)$.

It remains to prove (\ref{sumrule}) for $j=\ell$, for which one has to calculate $-\sum_kf_k^2(a)/(\omega_k+\omega_j)^2=\linebreak2\partial_\omega[\omega\tilde{G}^{\rm R}(a,a;\omega)]_{\omega=-\omega_j}$. To this end, define $f(x,\omega)$ ($g(y,\omega)$) as the solution of the homogeneous wave equation (\ref{def-QNM}) (upon the substitution $\omega_j\mapsto\omega$) satisfying the first (second) of the boundary conditions (\ref{f-bc}). This allows one to write
\beq
  \tilde{G}^{\rm R}(x{<}y;\omega)=
    \frac{f(x,\omega)\hp g(y,\omega)}{W(\omega)}\;,
\eeql{GfgW}
where one can choose
\beq
  f(x,\omega)=f(x,-\omega)\;,
\eeql{fsymm}
and where $W=fg'-gf'$ is the position-independent Wronskian of the functions $f$ and $g$\cite{comp1}. Together with the OWC for $f(x,\omega_j)=f_j(x)$ and $g$, (\ref{GfgW}) yields
\beq
  2\partial_\omega[\omega\tilde{G}^{\rm R}(a,a;\omega)]_{\omega=-\omega_j}=
  \frac{i\omega_j\partial_\omega f(a,\omega_j)+if_j(a)
        -\partial_\omega f'(a^+,\omega_j)}{2\omega_jf_j(a)}\;.
\eeql{dGdomega}
{\em If\/} $\omega_j$ were a double QNM pole\cite{jordan}, i.e., if $f$ satisfied the OWC up to and including ${\cal O}(\omega-\omega_j)$, the numerator would vanish. However, we assume simple poles throughout the main text (cf.\ Appendix~\ref{JBpath}), so further evaluation is needed. Solve $[\partial_x^2+\rho\omega^2]\partial_\omega f|_{\omega_j}=-2\omega_j\rho f_j$ by the variation-of-constant Ansatz $\partial_\omega f(x,\omega_j)=f_j(x)\hp h_j(x)$, leading to $h_j'(x)f_j^2(x)=-2\omega_j\int_0^xdy\,\rho(y)f_j^2(y)\Rightarrow h_j'(a^+)=i-(\fj,\fj)/f_j^2(a)\Rightarrow i\omega_j\partial_\omega f(a,\omega_j)+if_j(a)-\partial_\omega f'(a^+,\omega_j)=(\fj,\fj)/f_j(a)$; given our observation that (\ref{dGdomega}) vanishes for a double pole one could have expected an answer $\propto(\fj,\fj)$\cite{jordan}, which in the present case of course equals $2\omega_j$. Substituting back into (\ref{dGdomega}), one finds
\beq
  2\partial_\omega[\omega\tilde{G}^{\rm R}(a,a;\omega)]_{\omega=-\omega_j}=
  \frac{1}{f_j^2(a)},
\eeql{dGdom2}
completing the proof of the sum rule (\ref{sumrule}) and therefore of (\ref{complsq-a}).

\section{One-component forms}
\label{one-comp}

Given previous experience with the canonical approach\cite{sq}, one would expect that the cavity functional and effective action also have a diagonal form (cf.\ Appendix~\ref{diagApp}) if the source couples only to the first field component $\phi$. In this section we thus temporarily set the first source component to zero: $\chi=0$, and consider $S\{\hat{\chi}\}$. For $\chi=0$, the sum rule (\ref{sumrule}) implies that the $b_{jm}$ satisfy
\beq
  \sum_k\frac{f_j(a)f_k(a)}{\omega_j+\omega_k}\hp b_{km}=-ib_{jm}\;,
\eeql{bsum}
which at once yields
\beq
  S\{\hat{\chi}\}=\exp\Biggl\{T\sum_{jm}b_{jm}b_{j,-m}
  \frac{\omega_j}{\omega_j-i|\nu_m|}\Biggr\}\;.
\eeql{S1comp}
However, this form should be used with caution: the $f_j$ are overcomplete for the one-component expansion of $\hat{\chi}$, and consequently the $b_{jm}$ are now no longer independent, as exemplified by (\ref{bsum}). It might thus be more convenient to write out the projection formula $b_{jm}=(i/2\omega_j)\int_0^{a^+}\!dx\,f_j\hat{\chi}_m$, i.e.,
\bea
  S\{\hat{\chi}\}&=&\exp\Biggl\{-\frac{T}{4}\sum_{jm}\int_0^{a^+}\!\!dx\hp dy\,
  \hat{\chi}_m(x)\hat{\chi}_{-m}(y)
  \frac{f_j(x)f_j(y)}{\omega_j(\omega_j-i|\nu_m|)}\Biggr\}\label{S-chi}\\
  &=&\exp\Biggl\{\frac{T}{2}\sum_{m}\int_0^{a^+}\!\!dx\hp dy\,
  \hat{\chi}_m(x)\hat{\chi}_{-m}(y)\hp\tilde{\cal G}(x,y;\nu_m)\Biggr\}\;.
\eeal{S-G}
Eq.~(\ref{S-G}) looks very simple, yet this form (in particular the representation (\ref{S-chi}) of $\tilde{\cal G}$) would have been difficult to derive from (\ref{eff-act}) without using the power of the two-component QNM expansion in the intermediate steps, cf.\ below (\ref{act1comp}). Using the representation (\ref{S-chi}) for $S\{\hat{\chi}\}=\left<{\cal T}_\tau\exp\left\{i\int_0^\beta\!d\tau\int_0^{a^+}\!dx\,\hat{\chi}(x,\tau)\hp\phi(x,\tau)\right\}\right>$, $\phi$--$\phi$ correlators can be found by functional differentiation with respect to $\hat{\chi}$, without the problems associated with (\ref{S1comp}).

One can also obtain a diagonal representation for the effective cavity action. This might for instance be useful in the numerical study of an extended, interacting, model; cf.\ the discussion in Ref.\cite{sq}, Sections VI and~VII. For $\chi=0$, one can integrate out $\hat{\phi}$ and study the configuration-space path integral
\bea
  S\{\hat{\chi}\}&=&\int\frac{{\cal D}\phi_{\rm c}}{Z}
  \exp\Biggl\{T\sum_m\Biggl[\int_0^{a^+}\!\!dx\left(
  i\phi_m\hat{\chi}_{-m}-\frac{\rho\nu_m^2}{2}|\phi_m|^2-\frac{1}{2}|\phi_m'|^2
  \right)-\frac{|\nu_m|}{2}|\phi_m(a)|^2\Biggr]\Biggr\}\nonumber\\
  &=&\int\frac{{\cal D}\phi_{\rm c}}{Z}\exp\Biggl\{T\sum_m\Biggl[
    \sum_j2\omega_ja_{jm}b_{j,-m}-\sum_{jk}a_{jm}a_{k,-m}\nonumber\\
  &&\hphantom{\int\frac{{\cal D}\phi_{\rm c}}{Z}\exp\Biggl\{T\sum_m\Biggl[}
    \times\!\left(\frac{\nu_m^2+\omega_j^2}{2}\delta_{jk}
    +i\frac{(\omega_j-i|\nu_m|)(\omega_k-i|\nu_m|)}{2(\omega_j+\omega_k)}
    f_j(a)f_k(a)\right)\Biggr]\Biggr\}\,.
\eea
The $a_{jm}$ are now of course given by the projection
\bea
  a_{jm}&=&\frac{1}{2}\int_0^{a^+}\!\!dx\,\rho\phi_mf_j
    +\frac{i}{2\omega_j}\phi_m(a)f_j(a)\nonumber\\
  &=&\frac{1}{2\omega_j^2}\int_0^{a^+}\!\!dx\,\phi_m'f_j'\;,
\eea
implying
\beq
  \sum_k\frac{\omega_kf_j(a)f_k(a)}{\omega_j+\omega_k}\hp a_{km}=
  -i\omega_ja_{jm}\;.
\eeq
By also using $\sum_jf_j^2(a)/\omega_j^2=2a$ (see (\ref{Gtilde0})) and the representation $\phi_m(a)\phi_{-m}(a)=-i(\vec{\phi}_m,\vec{\phi}_{-m})=-2i\sum_j\omega_ja_{jm}a_{j,-m}$, the one-component action can be written as
\beq
  S\{\hat{\chi}\}=\int\frac{{\cal D}\phi_{\rm c}}{Z}
  \exp\biggl\{T\sum_{jm}a_{jm}\left[2\omega_jb_{j,-m}-a_{j,-m}
  \left(\nu_m^2+\omega_j^2-i|\nu_m|\omega_j-ia\nu_m^2\omega_j
  \right)\right]\biggr\}\;.
\eeql{act1comp}
Like for the one-component functional (\ref{S1comp}), one has the caveat that the $a_{jm}$ now no longer are independent degrees of freedom (as, apart from the constraint $a_{jm}^\phs=a_{-j,-m}^*$, they were in Section~\ref{cavQNM}). Therefore, direct evaluation of (\ref{act1comp}) is difficult, and the best route to (\ref{S1comp}) is probably via the auxiliary field $\hat{\phi}$ as has been done in the preceding. Again it might be more convenient to write out the expansion coefficients, viz.,
\bea
  S\{\hat{\chi}\}=\int\frac{{\cal D}\phi_{\rm c}}{Z}\exp\Biggl\{T\sum_m\Biggl[
  &&i\int_0^{a^+}\!\!dx\,\phi_m\hat{\chi}_{-m}-\frac{1}{4}
  \int_0^{a^+}\!\!dx\hp dy\,\phi_m'(x)\phi_{-m}'(y)\nonumber\\
  &&\times\sum_j\frac{\nu_m^2+\omega_j^2-i|\nu_m|\omega_j-ia\nu_m^2\omega_j}
  {\omega_j^4}f_j'(x)f_j'(y)\Biggr]\Biggr\}\;.
\eeal{act-phi}

\section{Discussion}
\label{discuss}

By carrying out the path-integral quantization of the open system (\ref{wave-eq})--(\ref{bc-right}), we have met the challenge set out at the end of Ref.\cite{sq}. This leaves us facing the second challenge mentioned there: the inclusion of matter in the system's Hamiltonian or action\cite{partII}. Assuming the interaction to be confined to the cavity, the elimination procedure of Section~\ref{elim} goes through unmodified and one simply has an extra term in the effective-action exponent of (\ref{eff-act}). It has already been mentioned in the Introduction that the analysis of Section~\ref{cav-int} consists of two steps. The first of these, expansion of the effective cavity action with respect to the QNM basis, certainly goes through for any form in the interaction since the QNMs are complete. One obtains an appropriate generalization of (\ref{S-QNM}), and the advantages of a discrete basis now apply to any qualitative, approximate or numerical analysis. While the second step of closed-form evaluation will not be possible any more in general, by substituting $a_{jm}\mapsto\partial/\partial(2\omega_jTb_{j,-m})$  in the interaction term one has the standard representation of the interacting $S\{\vec{\chi}\}$ as a functional of the free $S_0\{\vec{\chi}\}$\cite{ryder}, the latter being given by (\ref{Sres}). For instance, if one incorporates into the cavity action a toy nonlinear term ${\cal S}_{\rm int}=\int_0^\beta\hn d\tau\int_0^{a^+}\hn dx\,\lambda(x)\hp\phi^4(x,\tau)$ (so that the integrand of (\ref{pathint}) acquires an extra factor $\exp\{-{\cal S}_{\rm int}\}$), one can write
\beq
  S\{\vec{\chi}\}=
  Z^{-1}\exp\Biggl\{-\beta\!\!\sum_{m_1\ldots m_4}\delta_{m_1+\cdots+m_4}
  \sum_{j_1\ldots j_4}\lambda_{j_1\ldots j_4}\prod_{i=1}^4
  \frac{1}{2\omega_{j_i}}\frac{\partial}{\partial b_{j_im_i}}\Biggr\}
  S_0\{\vec{\chi}\}\;,
\eeq
with the coupling $\lambda_{j_1\ldots j_4}=\int_0^{a^+}\hn dx\,\lambda(x)f_{j_1}(x)\cdots f_{j_4}(x)$. This formal representation is a suitable starting point for perturbation theory.

It should be noted with some modesty that the final result (\ref{Sres}) could have been known in advance even without employing results of the canonical quantization of the system (cf.\ \cite{sq}, Section~III). Namely, the kernel (\ref{calG}) is the analytic continuation of $\tilde{G}^{\rm R}_{jk}(\omega)$ by (\ref{GtildeGR}). In their turn, the $G^{\rm R}_{jk}(t)$ are the unique QNM expansion coefficients of the tensor correlator ${\sf G}^{\rm R}(x,y;t)=-i\theta(t)\hp\langle[\vec{\phi}(x,t),\vec{\phi}(y)]\rangle$, all components of which follow by trivial time differentiations from its $(1,1)$ component (cf.\ (\ref{Gtensor})), the latter merely being the classical Green's function (\ref{greenfie}); see (\kern-1.6mm\cite{sq}, Section~VI~B and Appendix~C)\cite{tensorF}. Yet, the present explicit evaluation of the path integral is of considerable interest, since so few of them can be done in closed form unless they trivially decompose into a product of ordinary integrals over normal modes. Also, carrying out the calculation has uncovered new results on QNMs---the identity (\ref{sumrule}) and those of Appendix~\ref{sumApp}---which are useful already on the classical level.

\section*{Acknowledgment}

In the work's preliminary stage I benefited from a discussion with H. Dekker. K. Young is gratefully acknowledged for discussions, and for numerous comments on the manuscript. This work is supported in part by the Hong Kong Research Grants Council (Grant no.\ 452/95P).

\appendix

\section{Jordan-block path integral}
\label{JBpath}

\subsection{Evaluation of the integral}

The results in the main text have been derived under the assumption that all poles in $\tilde{G}^{\rm R}(\omega)$ are simple, as is for instance obvious by considering the Fourier transform of (\ref{greenfie}). Here we consider the general case, and follow Ref.\cite{jordan} throughout\cite{Gsign}. For each QNM pole in $\tilde{G}^{\rm R}(\omega)$ of order $M_j$ at $\omega=\omega_j$, one introduces
\beq
  f_{j,n}(x)=\frac{\partial_\omega^n|_{\omega_j}f(x,\omega)}{n!}
\eeql{fjn}
for $0\le n\le M_j-1$, with $f(x,\omega)$ defined above (\ref{GfgW}). The associated momenta read
\beq
  \hat{f}_{j,n}=-i\rho[\omega_jf_{j,n}+f_{j,n-1}]
\eeql{fjnhat}
($f_{j,n}\equiv0$ for $n\le-1$), so that $\vec{f}_{\!j,0}=\fj$ is the QNM eigenvector, which together with $\{\vec{f}_{\!j,n}\}_{n=1}^{M_j-1}$ spans a so-called Jordan block of the Hamiltonian (\ref{harray}).

One of the main results of\cite{jordan} can now be stated as follows. Provided that one chooses the functions $f$ and $g$ such that, for all $j$,
\beq
  g(x,\omega)=f(x,\omega)+{\cal O}[(\omega-\omega_j)^{M_j}]
\eeql{gchoice}
and
\beq
  W(\omega)=2\omega_j(\omega-\omega_j)^{M_j}+{\cal O}[(\omega-\omega_j)^{2M_j}]
\eeql{Wchoice}
(where we draw attention to the orders of the error terms), which is readily achieved, one has the biorthogonality relation
\beq
  (\vec{f}_{\!j,n},\vec{f}_{\!k,r})=2\omega_j\delta_{jk}\delta_{n+r,M_j-1}
\eeql{bior-eq}
for $0\le n\le M_j-1$, $0\le r\le M_k-1$.

Substituting the expansions $\vec{\phi}_m=\sum_j\sum_{n=0}^{M_j-1}a_{j,n;m}\vec{f}_{\!j,n}$ and $\vec{\chi}_m=\sum_j\sum_{n=0}^{M_j-1}b_{j,n;m}\vec{f}_{\!j,n}$ into (\ref{eff-act}), the term involving $f_{j,n}'$ can again be dealt with using an integration by parts, supplemented by the QWC $f_{j,n}'(a^+)+\hat{f}_{j,n}(a^+)=0$ and the wave equation $[\partial_x^2+\rho\omega_j^2]f_{j,n}=-\rho[2\omega_jf_{j,n-1}+f_{j,n-2}]$. However, integrals $\int_0^{a^+}\!dx\,\rho f_{j,n}f_{k,r}$ in general cannot be reduced to surface terms by a single application of (\ref{bior-eq}) as in (\ref{ortho2}), because of the second term in (\ref{fjnhat}). Rather, iteration leads to
\bea
  \int_0^{a^+}\!\!dx\,\rho f_{j,n}f_{k,r}&=&
    \frac{\delta_{jk}\theta(n+r+\frac{3}{2}-M_j)}{(-\omega_j)^{n+r+1-M_j}}
    +i\sum_{p=0}^n\sum_{q=0}^r\left(\begin{array}{c} p+q \\ p \end{array}\right)
    \frac{f_{j,n-p}(a)f_{k,r-q}(a)}{(-\omega_j-\omega_k)^{p+q+1}}\label{Ijnkr}\\
  &=&\frac{\delta_{jk}\theta(n+r+\frac{3}{2}-M_j)}{(-\omega_j)^{n+r+1-M_j}}
    -i\frac{\partial_\omega^n|_{\omega_j}}{n!}
    \frac{\partial_\mu^r|_{\omega_k}}{r!}
    \frac{f(a,\omega)f(a,\mu)}{\omega+\mu}\;,
\eeal{Ijnkr2}
where the equivalence of the second line to the first follows from (\ref{fjn}). Comparatively compact forms like (\ref{Ijnkr2}) will be essential in the calculation below. Comparing the second (non-diagonal) terms on the respective r.h.s.'s of (\ref{ortho2}) and (\ref{Ijnkr2}), the latter is seen to be a differentiated version of the former. However, it should be borne in mind that (\ref{Ijnkr2}) only holds for the special choice of $f$ satisfying (\ref{gchoice}) and (\ref{Wchoice}).

Returning to (\ref{eff-act}), the above yields
\bea
  S\{\vec{\chi}\}=\int\frac{{\cal D}\vec{\phi}_{\rm c}}{Z}\,\exp\Biggl\{
  T\sum_m\Biggl[&&-\sum_{jk}
  \sum_{n=0}^{M_{j\vphantom{k}}-1}\sum_{r=0}^{M_{k\vphantom{j}}-1}
  a_{j,n;m}a_{k,r;-m}\tilde{\cal S}_{j,n;k,r}(\nu_m)\nonumber\\
  &&{}+\sum_j\sum_{n=0}^{M_j-1}
  2\omega_ja_{j,n;m}b_{j,M_j-1-n;-m}\Biggr]\Biggr\}\;,
\eeal{S-calS}
with the action matrix
\beq
  \tilde{\cal S}_{j,n;k,r}(\nu_m)=\frac{\partial_\omega^n|_{\omega_j}}{n!}
  \frac{\partial_\mu^r|_{\omega_k}}{r!}f(a,\omega)f(a,\mu)
  \frac{\nu_m(\mu\theta(m)-\omega\theta(-m))+i\omega\mu}{\omega+\mu}\;.
\eeql{Sjnkr}
It is seen that the first term of (\ref{Ijnkr2}) cancels in $\tilde{\cal S}$, cf.\ the remark below (\ref{abar-a}).

We claim that the result of (\ref{S-calS}) reads
\beq
  S\{\vec{\chi}\}=\exp\biggl\{-2T\sum_{jkm}
  \sum_{n=0}^{M_{j\vphantom{k}}-1}\sum_{r=0}^{M_{k\vphantom{j}}-1}
  \omega_j\omega_kb_{j,n;m}b_{k,r;-m}
  \tilde{\cal G}_{j,M_j-1-n;k,M_k-1-r}(-\nu_m)\biggr\}\;,
\eeql{SJB}
in terms of the Jordan-block temperature Green's function
\beq
  \tilde{\cal G}_{j,n;k,r}(\nu_m)=
  \frac{\partial_\lambda^{M_j-1-n}|_{\omega_j}}{(M_j-1-n)!}
  \frac{\partial_\mu^{M_k-1-r}|_{\omega_k}}{(M_k-1-r)!}
  \frac{f(a,\lambda)f(a,\mu)}{2\omega_j\omega_k(\lambda+\mu)}\left\{
  \frac{\theta(m)\lambda}{i\lambda+\nu_m}
  +\frac{\theta(-m)\mu}{i\mu-\nu_m}\right\}\;.
\eeql{Gjnkr}

To verify this claim, it suffices to show that\cite{domain}
\beq
  \sum_k\sum_{r=0}^{M_k-1}
  \tilde{\cal S}_{j,n;k,r}(\nu_m)\hp\tilde{\cal G}_{\ell,u;k,r}(\nu_m)=
  -\frac{1}{2}\delta_{j\ell}\delta_{nu}=
  \sum_k\sum_{r=0}^{M_k-1}
  \tilde{\cal S}_{k,r;j,n}(\nu_m)\hp\tilde{\cal G}_{k,r;\ell,u}(\nu_m)\;,
\eeql{SGprod}
where in view of the simple symmetry $\tilde{\cal S}_{j,n;k,r}(\nu_m)=\tilde{\cal S}_{k,r;j,n}(-\nu_m)$ and likewise for $\tilde{\cal G}$, the second of these relations is equivalent to the first. Therefore, let us evaluate the l.h.s.\ of (\ref{SGprod}), first assuming $\nu_m>0$. Substituting (\ref{Sjnkr}) and (\ref{Gjnkr}), the sum over $r$ in (\ref{SGprod}) can be performed using the product rule for differentiation, and one obtains
\bea
  \sum_k&&\sum_{r=0}^{M_k-1}
  \tilde{\cal S}_{j,n;k,r}(\nu_m)\hp\tilde{\cal G}_{\ell,u;k,r}(\nu_m)
  =\nonumber\\
  &&\frac{\partial_\omega^n|_{\omega_j}}{n!}
  \frac{\partial_\lambda^{M_\ell-1-u}|_{\omega_\ell}}{(M_\ell-1-u)!}
  \frac{\nu_m+i\omega}{\nu_m+i\lambda}\frac{\lambda}{\omega_\ell}
  f(a,\omega)f(a,\lambda)
  \sum_k\frac{\partial_\mu^{M_k-1}|_{\omega_k}}{(M_k-1)!}
  \frac{\mu f(a,\mu)^2}{2\omega_k(\mu+\omega)(\mu+\lambda)}\;.
\eeal{SG2}
Carrying out the partial-fraction expansion
\beq
  \frac{\mu}{(\mu+\omega)(\mu+\lambda)}=\frac{1}{\omega-\lambda}
  \left[\frac{\omega}{\mu+\omega}-\frac{\lambda}{\mu+\lambda}\right]\;,
\eeql{parfrac}
one recognizes in (\ref{SG2}) the QNM expansion of the Green's function\cite{jordan}
\beq
  \tilde{G}^{\rm R}(x,y;\zeta)=
  \sum_k\frac{\partial_\mu^{M_k-1}|_{\omega_k}}{(M_k-1)!}
  \frac{f(x,\mu)f(y,\mu)}{2\omega_k(\zeta-\mu)}\;.
\eeql{GJB}
One can now replace (\ref{GJB}) by the alternate form (\ref{GfgW}) (which holds independently of the pole structure of $\tilde{G}^{\rm R}$), and in the latter use (\ref{fsymm}) to write
\beq
  \tilde{G}^{\rm R}(x,a;-\omega)=
  -\frac{f(x,\omega)}{f'(a^+,\omega)+i\omega f(a,\omega)}\;.
\eeq
The denominator can be simplified by
\bea
  i\omega f(a,\omega)-f'(a^+,\omega)&=&
  \frac{2\omega_j(\omega-\omega_j)^{M_j}}{g(a,\omega)}
  +{\cal O}[(\omega-\omega_j)^{2M_j}]\nonumber\\
  &=&\frac{2\omega_j(\omega-\omega_j)^{M_j}}{f(a,\omega)}
  +{\cal O}[(\omega-\omega_j)^{2M_j}]\;,
\eea
where we subsequently used (\ref{Wchoice}) and (\ref{gchoice}), leading to
\beq
  \omega\tilde{G}^{\rm R}(x,a;-\omega)=\frac{if(x,\omega)}{2f(a,\omega)}
  +\frac{\omega_jf(x,\omega)\hp(\omega-\omega_j)^{M_j}}{2\omega f^3(a,\omega)}
  +{\cal O}[(\omega-\omega_j)^{2M_j}]\;,
\eeql{Gminmu}
which is seen to generalize both (\ref{GminOj}) and (\ref{dGdom2}) simultaneously. The last (higher-order) term of (\ref{Gminmu}) is readily verified not to contribute in the final result (\ref{fin-res}) below (however, for this it is essential that it be at least of the order indicated), and from now on will not be written explicitly. Substitute (\ref{Gminmu}) (with $x\mapsto a$, and with $(\omega,j)\mapsto(\lambda,\ell)$ in the second term resulting from (\ref{parfrac})) into the upshot of (\ref{SG2}), one arrives at
\bea
  \sum_k&&\sum_{r=0}^{M_k-1}
  \tilde{\cal S}_{j,n;k,r}(\nu_m)\hp\tilde{\cal G}_{\ell,u;k,r}(\nu_m)
  =\nonumber\\
  &&\frac{\partial_\omega^n|_{\omega_j}}{n!}
  \frac{\partial_\lambda^{M_\ell-1-u}|_{\omega_\ell}}{(M_\ell-1-u)!}
  \frac{\nu_m+i\omega}{\nu_m+i\lambda}\frac{\lambda}{\omega_\ell}
  \frac{f(a,\omega)f(a,\lambda)}{\omega-\lambda}\left[
  \frac{\omega_\ell\hp(\lambda-\omega_\ell)^{M_\ell}}{2\lambda f^2(a,\lambda)}
  -\frac{\omega_j\hp(\omega-\omega_j)^{M_j}}{2\omega f^2(a,\omega)}\right]\,.
\eeal{SG3}
Since $n\le M_j-1$ and $u\ge0$, one immediately sees that this vanishes if $j\neq\ell$, in which case the denominator $\omega-\lambda$ does not become singular. If $j=\ell$, in the first term of (\ref{SG3}) one can use $(\nu_m+i\omega)f(a,\omega)/[(\nu_m+i\lambda)f(a,\lambda)]=1+{\cal O}(\omega-\lambda)$, where the higher-order term does not contribute because it cancels the singular denominator $\omega-\lambda$, upon which $(\lambda-\omega_j)^{M_j}$ yields zero in the final differentiation. The second term of (\ref{SG3}) is handled analogously, and one is left with
\bea
  \sum_k\sum_{r=0}^{M_k-1}
  \tilde{\cal S}_{j,n;k,r}(\nu_m)\hp\tilde{\cal G}_{j,u;k,r}(\nu_m)
  &=&\frac{\partial_\omega^n|_{\omega_j}}{n!}
  \frac{\partial_\lambda^{M_j-1-u}|_{\omega_j}}{(M_j-1-u)!}
  \frac{(\lambda-\omega_j)^{M_j}-(\omega-\omega_j)^{M_j}}
  {2(\omega-\lambda)}\nonumber\\
  &=&-\frac{1}{2}\frac{\partial_\omega^n|_{\omega_j}}{n!}
  \frac{\partial_\lambda^{M_j-1-u}|_{\omega_j}}{(M_j-1-u)!}
  \sum_{p=0}^{M_j-1}(\omega-\omega_j)^p(\lambda-\omega_j)^{M_j-1-p}\nonumber\\
  &=&-\frac{1}{2}\delta_{nu}\;,
\eeal{fin-res}
which proves the claim made above (\ref{SJB}) for the case $\nu_m>0$. The case $\nu_m\le0$ is fully analogous but slightly simpler, since the factors involving $\nu_m$ cancel from the outset.

\subsection{Comparison with canonical quantization}

Let us finally make a brief comparison with the operator version of the quantum theory, for an independent check on the above algebra and for a better understanding of the unusual quantum excitations corresponding to the higher-order poles in the correlation functions. It turns out that the generalization of Refs.\cite{KL,sq} is straightforward. Expanding the operator field $\vec{\phi}(t)=\sum_j\sum_{n=0}^{M_j-1}a_{j,n}(t)\vec{f}_{\!j,n}$, the operator QNM expansion coefficients $a_{j,n}$ satisfy the coupled (for $M_j\ge2$) system of quantum Langevin equations
\beq
  \dot{a}_{j,n}+i\omega_ja_{j,n}+ia_{j,n+1}=
  \frac{i}{2\omega_j}f_{j,M_j-1-n}(a)\hp b(t)\;,
\eeql{langevin}
for $0\le n\le M_j-1$ (with $a_{j,M_j}\equiv0$), with the environmental driving force $b(t)=2\hat{\phi}_{\rm in}(a+t)\equiv\linebreak\phi'(a^+,t)+\hat{\phi}(a^+,t)$ (Eq.~(4.2) in\cite{sq}), which is independent of the QNM pole structure. The hierarchy (\ref{langevin}) yields $\tilde{a}_{j,n}(\omega)=(i/2\omega_j)\tilde{f}_{j,M_j-1-n}(a,\omega)\hp\tilde{b}(\omega)$ in terms of the functions $\tilde{f}_{j,n}(x,\omega)\equiv\int_0^\infty\!dt\,e^{i\omega t}f_{j,n}(x,t)=\int_0^\infty\!dt\,e^{i\omega t}(n!)^{-1}\partial_\mu^n|_{\omega_j}[f(x,\mu)e^{-i\mu t}]=i\sum_{p=0}^nf_{j,n-p}(x)\times\linebreak(\omega-\omega_j)^{-p-1}=(i/n!)\partial_\mu^n|_{\omega_j}[f(x,\mu)/(\omega-\mu)]$ (cf.\cite{jordan} for $f_{j,n}(x,t)$). For the correlators this implies
\bea
  \tilde{F}_{j,n;k,r}(\omega)&\equiv&
  \left<\tilde{a}_{j,n}(\omega)\hp a_{k,r}\right>\nonumber\\
  &=&-\frac{\tilde{f}_{j,M_j-1-n}(a,\omega)\tilde{f}_{k,M_k-1-r}(a,-\omega)}
  {4\omega_j\omega_k}\hp\langle\tilde{b}(\omega)b\rangle\;,
\eeal{Fjnkr}
where the driving-force correlator reads $\langle\tilde{b}(\omega)b\rangle=2\omega/(1-e^{-\beta\omega})$ as for simple poles\linebreak(\kern-1.6mm\cite{sq}, Eq.~(6.3)).

The retarded propagator may be found from (\ref{Fjnkr}) as $\tilde{G}^{\rm R}_{j,n;k,r}(\omega)= \int\hn(d\omega'/2\pi)\times\linebreak(\omega-\omega'+i\epsilon)^{-1}[\tilde{F}_{j,n;k,r}(\omega')-\tilde{F}_{k,r;j,n}(-\omega')]$ (cf.\cite{sq}, Eq.~(7.2)), leading to
\beq
  \tilde{G}^{\rm R}_{j,n;k,r}(\omega)=\frac{i}{2\omega_j\omega_k}
  \frac{\partial_\lambda^{M_j-1-n}|_{\omega_j}}{(M_j-1-n)!}
  \frac{\partial_\mu^{M_k-1-r}|_{\omega_k}}{(M_k-1-r)!}
  \frac{\lambda f(a,\lambda)f(a,\mu)}{(\omega-\lambda)(\lambda+\mu)}\;.
\eeql{GRjnkr}
These are the QNM expansion coefficients of the tensor correlator ${\sf G}^{\rm R}(x,y;t)\equiv-i\theta(t)\*\hp\langle[\vec{\phi}(x,t),\vec{\phi}(y)]\rangle=\sum_{jk}\sum_{n=0}^{M_j-1}\sum_{r=0}^{M_k-1}G^{\rm R}_{j,n;k,r}(t)\vec{f}_{\!j,n}(x)\vec{f}_{\!k,r}(y)$. Since the latter is related to the classical Green's function by
\beq
  \tilde{\sf G}^{\rm R}(x,y;\omega)=
  \left(\begin{array}{cc} 1 & i\omega\rho(y) \\
        -i\omega\rho(x) & \omega^2\rho(x)\rho(y) \end{array}\right)
  \tilde{G}^{\rm R}(x,y;\omega)
  -\left(\begin{array}{cc} 0 & 0 \\ 0 & \rho(x)\hp\delta(x-y) \end{array}\right)
\eeql{Gtensor}
(cf.\cite{sq}, Eq.~(6.6); the second term in (\ref{Gtensor}) comes from time differentiation of the step function in the definition of ${\sf G}^{\rm R}$), (\ref{GRjnkr}) can also be obtained directly from the classical theory by substituting (\ref{GJB}) into (\ref{Gtensor}) and subsequently applying the tensor-product projection formula (i.e., Appendix~C of\cite{sq}, generalized to the present case of nontrivial Jordan blocks, which merely involves using (\ref{fjnhat}) for the second components and keeping track of the intra-block indices as stipulated by (\ref{bior-eq})). Conversely, using the sum rules implicit in (\ref{Gminmu}), the expression $\tilde{G}^{\rm R}(x,y;\omega)=\sum_{jk}\sum_{n=0}^{M_j-1}\sum_{r=0}^{M_k-1}\tilde{G}^{\rm R}_{j,n;k,r}(\omega)f_{j,n}(x)f_{k,r}(y)$ is readily reduced to the form (\ref{GJB}).

Since inspection shows that $\tilde{\cal G}$ as in (\ref{Gjnkr}) and $\tilde{G}^{\rm R}$ as in (\ref{GRjnkr}) are related as in (\ref{GtildeGR}), the path integral and canonical approaches to the quantum theory agree for arbitrary QNM pole configurations.

\section{QNM sum rules}
\label{sumApp}

In this appendix we give an overview of some sum rules which have been used in the above, as well as more direct proofs of some others which can be read off by comparing the various forms for $S\{\hat{\chi}\}$ in the main text. A central role is played by the WKB approximation to the Green's function\cite{comp1},
\beq
  \tilde{G}^{\rm R}(x\le y\le a;\omega)\approx-\frac{\sin(\omega T(x))
  \left[e^{-i\omega T(y,a)}+R(\omega)e^{i\omega T(y,a)}\right]}
  {\omega\sqrt{n(x)n(y)}\left[e^{-i\omega T}+R(\omega)e^{i\omega T}\right]}\;,
\eeql{GWKB}
where $n\equiv\sqrt{\rho}$, and with the geometric-optics transit time $T(x,y)=\int_x^y\!ds\,n(s)$, $T(x)\equiv T(0,x)$, $T\equiv T(a)$. For $x\le a\le y$ one has (exactly) $\tilde{G}^{\rm R}(x,y;\omega)=\tilde{G}^{\rm R}(x,a;\omega)e^{i\omega(y-a)}$. If the discontinuity at $x=a$ is at least a step, the reflection amplitude is
\beq
  R(\omega)=\frac{n^--1+i\mu\omega}{n^-+1-i\mu\omega}\;;
\eeql{refl}
for weaker discontinuities we write $R(\omega)\sim R_p/\omega^p$ for some integer $p\ge1$.

Standard wisdom states that $G^{\rm R}(t{=}0)=0$, but in fact this is only true distributionally, and pointwise for $x\neq y$. Examining the spreading-plateau solution for $G^{\rm R}(x,y;t)$ for small positive $t$ (where $\rho(x{\approx}y)$ can be taken constant), one finds that $G^{\rm R}(x,x;0^+)=-1/2n(x)$ if $n$ is continuous at the point $x$; at $x=a$ one finds instead $G^{\rm R}(a,a;0^+)=-1/(n^-+1)$ ($n^-\equiv n(a^-)$) or $G^{\rm R}(a,a;0^+)=0$ if $\mu>0$ (other possible points of discontinuity inside the cavity work analogously). Since certainly $G^{\rm R}(t{=}0^-)=0$, by the general theory of Fourier integrals one expects $G^{\rm R}(t{=}0)$ to be the average of these values, i.e., $G^{\rm R}(x,x;0)=-1/2(n(x^-)+n(x^+))$; this can be verified by performing the countour integration of $\tilde{G}^{\rm R}$ in the upper half plane, where only the large semicircle contributes. Subsequently, one can integrate in the lower half plane, which upon comparison yields the sum over QNM residues.

Define
\beq
  s(x,y;t)=\sum_j\frac{f_j(x)f_j(y)}{\omega_j}e^{-i\omega_jt}\;;
\eeql{def-s}
since the terms of this sum only tend to zero if $t>T(x)+T(y)-2T/p$\cite{comp2}, and since we only want to consider (\ref{def-s}) for $t=0$ and $t=0^+$ but for all $x,y$ in the cavity, the analysis will be restricted to $p\le1$. If $t=0$, (\ref{def-s}) in general has to be understood as a principal value $\lim_{M\rightarrow\infty}\sum_{j=-M}^M$. Choosing $x\le y$, one has the following cases.
\beq\begin{array}{ll}
  x=0:& s=0\qquad\mbox{(all terms zero)}\\
  0<x<y\le a:& s=0\\
  0<x<a\le y<a+T(x,a):& s=0\\[2mm]
  0<x=y<a:& s=\left\{\begin{array}{ll}0, & t=0 \\
    \ds-\frac{i}{n(x)}, & t=0^+\end{array}\right.\\[7.5mm]
  x=y=a,\;t=0:& s=\left\{\begin{array}{ll}0, & \mu>0 \\
    \ds\frac{2i}{(n^-)^2-1}, & \mbox{step} \\[4mm] \mbox{divergent}, & p=1
    \end{array}\right.\\[1cm]
  x=y=a,\;t=0^+:& s=\left\{\begin{array}{ll}0, & \mu>0 \\
    \ds-\frac{2i}{n^-+1}, & \mbox{step} \\[3mm] -i, & p=1
    \end{array}\right.\\[1cm]
  0<x<a,\;y=a+T(x,a),\;t=0:& s=\left\{\begin{array}{ll}0, & \mu>0 \\
    \ds\frac{i}{n^--1}\sqrt{\vphantom{\frac{n}{n}}\smash{\frac{n^-}{n(x)}}},
    & \mbox{step} \\[3mm] \mbox{divergent}, & p=1 \end{array}\right.\\[1cm]
  0<x<a,\;y=a+T(x,a),\;t=0^+:& s=0\\[2mm]
  0<x\le a,\;y=a+T(x,a)+\epsilon\;(\epsilon\downarrow0):&
    s=\left\{\begin{array}{ll}0, & \mu>0 \\
    \ds\frac{2i}{n^--1}\sqrt{\vphantom{\frac{n}{n}}\smash{\frac{n^-}{n(x)}}},
    & \mbox{step} \\[3mm] \mbox{divergent}, & p=1 \end{array}\right.
\end{array}\eeql{sumtable}
Some comments on this table: note that results for $\mu>0$ follow from those for a step by formally setting $n^-\rightarrow\infty$, and those for $p=1$ follow by letting $n^-\rightarrow1$. The result for $x=y=a$, $t=0^+$, $p=1$ also follows from the one for $0<x=y<a$, $t=0^+$, $p=1$ by letting $x\uparrow a$ (since $n^-=1$ if $p=1$). In this latter case, the factor $e^{-i\omega_j0^+}$ serves not only as an oscillating regulator, but also as a power-law damping. If there is at least a step at $x=a$, however, the sums with $t=0$ converge faster than the regulated ones, since in the absence of a factor $e^{-i\omega_jt}$ the leading ${\cal O}(j^{-1})$ tails in the summand for positive and negative $j$ cancel exactly. These results can be checked for the ``dielectric rod" $\rho(x{<}a)=\mbox{const}$\cite{sq}, where they become conventional Fourier series.

With (\ref{sumtable}) at our disposal, we can return to (\ref{offd-sum}). In the second term on the r.h.s., the summand has a denominator $\sim\omega_k^2$, so that one can indeed take $x\uparrow a$ in (\ref{GminOj}), as has been done in the main text. In the case of a step in $\rho(x)$ at $x=a$, (\ref{sumtable}) leads to the result for (\ref{offd-sum}) quoted below (\ref{GminOj}); for $\mu>0$, one has $\sum_kf_k^2(a)/(\omega_k+\omega_j)=i$ instead. For $p=1$ one finds that $\sum_kf_k^2(a)/(\omega_k+\omega_j)$ diverges, but this merely means that the partial-fraction expansion applied to (\ref{sumrule}) in that case is valid only term by term. In fact, one can readily verify that (\ref{sumrule}) itself remains convergent and valid also if $p=1$.

A different QNM sum is $\sum_jf_j(x)f_j(y)/\omega_j^2=-2\tilde{G}^{\rm R}(x,y;0)$ for $0\le x\le a$, $x\le y\le a+T(x,a)$ and $x\leftrightarrow y$. For $\omega=0$, the differential equation for $\tilde{G}^{\rm R}$ is trivial, and one finds
\beq
  \sum_j\frac{f_j(x)f_j(y)}{\omega_j^2}=2\min(x,y)\;.
\eeql{Gtilde0}
Operating on this with $\partial_x^2$ reproduces the familiar $\rho(x)\sum_jf_j(x)f_j(y)=2\delta(x-y)$ for\linebreak $0<x,y<a$ (more distributional sums are given below).

The form (\ref{act-phi}) for $S\{\hat{\chi}\}$ also prompts one to study $\sum_jf_j(x)f_j(y)/\omega_j^3$ and $\sum_jf_j(x)\*f_j(y)/\omega_j^4$; I restrict myself to $0\le x,y\le a$. In both cases one can operate with $\partial_x^2$ and $\partial_y^2$ termwise for $0<x,y<a$, and determine these derivatives from (\ref{sumtable}) and (\ref{Gtilde0}). Since furthermore the sums vanish for $x=0$ or $y=0$, one has
\bea
  \sum_j\frac{f_j(x)f_j(y)}{\omega_j^3}&=&C_3xy\nonumber\\
  \sum_j\frac{f_j(x)f_j(y)}{\omega_j^4}&=&
    C_4xy+2\int_0^{a^+}\!\!ds\,\rho(s)\min(s,x)\min(s,y)\;.
\eea
The constants follow from
\bea
  \partial_x\left.\sum_j\frac{f_j(x)f_j(y)}{\omega_j^3}\right|_{x=a^-}&=&C_3y
  \nonumber\\ &=&i\sum_j\frac{f_j(a)f_j(y)}{\omega_j^2}
  +\mu\sum_j\frac{f_j(a)f_j(y)}{\omega_j}\nonumber\\
  &=&2iy\qquad\Rightarrow\nonumber\\
  C_3&=&2i\;,
\eea
and
\bea
  \partial_x\left.\sum_j\frac{f_j(x)f_j(y)}{\omega_j^4}\right|_{x=a^-}
  &=&(C_4+2\mu)y \nonumber\\
  &=&i\sum_j\frac{f_j(a)f_j(y)}{\omega_j^3}
  +\mu\sum_j\frac{f_j(a)f_j(y)}{\omega_j^2}\nonumber\\
  &=&2(\mu-a)y\qquad\Rightarrow\nonumber\\
  C_4&=&-2a\;.
\eea

These classical sum rules combine nicely to yield some distributional identities. We already know $\rho(x)\sum_jf_j(x)f_j(y)=2\delta(x-y)$ if $0<x,y<a$, but what if $y\uparrow a$? This question often arises in the course of calculations. However, we know little about QNM sums outside $[0,a]$, and we want distributions to operate on functions defined in that interval only. Thus, we want to evaluate $q(x)=\sum_j\theta(a^+-x)\rho(x)f_j(x)f_j(a)$; note how writing the discontinuous (singular for $\mu>0$) $\theta\rho$ after the sum avoids the need to multiply this factor with the singular (discontinuous for $\mu>0$) $\sum_jf_j(x)f_j(a)$, so that as written $q(x)$ is well-defined. One evaluates
\bea
  q(x)&=&-\sum_j\theta(a^+-x)\hp d_x^2\frac{f_j(x)f_j(a)}{\omega_j^2}\nonumber\\
  &=&-d_x^2\sum_j\theta(a-x)\frac{f_j(x)f_j(a)}{\omega_j^2}
    +\sum_j(d_x\theta(a^+-x))\hp d_x\frac{f_j(x)f_j(a)}{\omega_j^2}\nonumber\\
    &&{}+d_x\sum_j(d_x\theta(a-x))\frac{f_j(x)f_j(a)}{\omega_j^2}\nonumber\\
  &=&-d_x^2(\theta(a-x)2x)-\sum_j\delta(x-a^+)\frac{f_j'(x)f_j(a)}{\omega_j^2}
    -d_x\sum_j\delta(x-a)\frac{f_j(x)f_j(a)}{\omega_j^2}\nonumber\\
  &=&2\delta(x-a)+2a\delta'(x-a)-i\delta(x-a)\sum_j\frac{f_j^2(a)}{\omega_j}
    -d_x(\delta(x-a)2a)\nonumber\\
  &=&\left\{\begin{array}{ll} 2\delta(x-a)\;, & \mu>0 \\
    \ds\left(2+\frac{2}{(n^-)^2-1}\right)\delta(x-a)
    =\frac{2(n^-)^2}{(n^-)^2-1}\delta(x-a)\;, & \mbox{step} \\[4mm]
    \mbox{divergent}\;, & p=1\;. \end{array}\right.
\eeal{dist-sum1}
The step leading to the one but last line is a bit dubious for $\mu>0$, since in the second term on the preceding line $\delta(x-a)$ multiplies a discontinuous $f_j'(x)$. However, for $\mu>0$ the sum rule (\ref{dist-sum1}) merely states that $\sum_jf_j^2(a)=2/\mu$, which can be verified without any distributional trickery by contour integration of $\omega\hp\tilde{G}^{\rm R}(a,a;\omega)$. Since $f_j(a)={\cal O}(j^{-1})$ if $\mu>0$ the sum converges quadratically, hence also $\sum_jf_j^2(a)e^{-i\omega_j0^+}=\sum_jf_j(a)f_j(a+\epsilon)\;(\epsilon\downarrow0)\;=2/\mu$. On the other hand, if $\mu>0$ and $x<a$ one has $\sum_jf_j(x)f_j(a)=0$.

Returning to $q(x)$, it is straightforward to repeat the calculation for a small positive $t$, leading to
\beq
  \sum_j\theta(a^+-x)\hp\rho(x)f_j(x)f_j(a)e^{-i\omega_j0^+}=
    \left\{\begin{array}{ll} 2\delta(x-a)\;, & \mu>0 \\
    \ds\frac{2n^-}{n^-+1}\delta(x-a)\;, & \mbox{step} \\[4mm]
    \delta(x-a)\;, & p=1\;. \end{array}\right.
\eeq
Comparing the outcome with the familiar case $y<a$, one sees that the presence of the $\theta$-function has no effect if $\mu>0$ (since the whole weight of the surface term in $\rho$ falls inside $[0,a^+)$). However, if $\rho(x)$ is continuous at $x=a$, the factor $\theta(a-x)$ halves the weight of the QNM sum, as could be expected. The case of a step in $\rho(x)$ at $x=a$ is intermediate between these two (for $n^->1$).

For reference I also give the ``unrestricted" form:
\bea
  \sum_j\rho(x)f_j(x)f_j(a)&=&-d_x^2\frac{f_j(x)f_j(a)}{\omega_j^2}\nonumber\\
  &=&\delta(x-a)\left[d_x\left.\frac{f_j(x)f_j(a)}{\omega_j^2}\right|_{x=a^-}
    -d_x\left.\frac{f_j(x)f_j(a)}{\omega_j^2}\right|_{x=a^+}\right]\nonumber\\
  &=&\delta(x-a)\left[2-i\frac{f_j(a)f_j(a+\epsilon)}{\omega_j}\right]
    \quad(\epsilon\downarrow0)\nonumber\\
  &=&\left\{\begin{array}{ll} 2\delta(x-a)\;, & \mu>0 \\
    \ds\frac{2n^-}{n^--1}\delta(x-a)\;, & \mbox{step} \\[4mm]
    \mbox{divergent}\;, & p=1\;. \end{array}\right.
\eea
However, this last sum rule is only valid up to a finite regular contribution for $x>a$, so it can only be used inside integrals $\int_0^{a^+}\!dx$.

Finally, I consider an object which is one step more singular, defining $r(x,y)=\sum_j\omega_j\theta(a^+-x)\hp\theta(a^+-y)\rho(x)\rho(y)f_j(x)f_j(y)$, which one for instance encounters upon substituting the projection formula for $a_j$ into $\hat{\phi}=\sum_ja_j\hat{f}_j$. For variation, I present the calculation using a test function:
\bea
  \sum_j\omega_j\int_0^{a^+}\!\!dx\hp dy\,&&\rho(x)\rho(y)f_j(x)f_j(y)\phi(x,y)
  \nonumber\\
  =&&-\sum_j\int_0^{a^+}\!\!dx\hp dy\,\rho(y)\frac{f_j''(x)f_j(y)}{\omega_j}\hp
  \phi(x,y)\nonumber\\
  =&&-\sum_j\int_0^{a^+}\!\!dy\,\rho(y)\frac{f_j'(a)f_j(y)}{\omega_j}\hp
    \phi(a,y)+\sum_j\int_0^{a^+}\!\!dx\hp dy\,\rho(y)
    \frac{f_j'(x)f_j(y)}{\omega_j}\hp \partial_x\phi(x,y)\nonumber\\
  =&&-i\sum_j\int_0^{a^+}\!\!dy\,\rho(y)f_j(a)f_j(y)\phi(a,y)
    +\sum_j\int_0^{a^+}\!\!dy\,\rho(y)\frac{f_j(a)f_j(y)}{\omega_j}\hp
    \partial_x\phi(a,y)\nonumber\\
    &&{}-\sum_j\int_0^{a^+}\!\!dx\hp dy\,\rho(y)
    \frac{f_j(x)f_j(y)}{\omega_j}\hp\partial_x^2\phi(x,y)\nonumber\\
  =&&-\frac{2i(n^-)^2}{(n^-)^2-1}\hp\phi(a,a)\qquad\Rightarrow\
\eeal{calc-rxy}
\beq
  r(x,y)=-\frac{2i(n^-)^2}{(n^-)^2-1}\hp\delta(x-a)\hp\delta(y-a)\;.
\eeql{rxy}
In the step leading to the third line of (\ref{calc-rxy}) we supposed our test function to satisfy $\phi(0,y)=0$; in the one but last line of (\ref{calc-rxy}), the second and third term vanish on behalf of (\ref{sumtable}) for $0<x<y\le a$.

Eq.~(\ref{rxy}) leads one to suspect that, for $\mu>0$, $\sum_j\omega_jf_j^2(a)=-2i/\mu^2$. To verify this, integrate $[\omega^2/(1+\omega^2\hn/\hn K^2)]\tilde{G}^{\rm R}(a,a;\omega)$ over the circle $|\omega|=R$. As $R\rightarrow\infty$ for fixed $K$, the contour integral vanishes. Now let $K\rightarrow\infty$; the QNM pole contributions can be evaluated for $K=\infty$, and the residues at $\omega=\pm iK$ can be evaluated in the WKB approximation. Combination gives the desired result.

With only a trivial variation, one obtains
\beq
  \sum_j\omega_j\theta(a^+-x)\hp\theta(a^+-y)\rho(x)\rho(y)f_j(x)f_j(y)
  e^{-i\omega_j0^+}=
  \left\{\!\!\begin{array}{ll} \ds -\frac{2in^-}{n^-+1}\hp
    \delta(x-a)\hp\delta(y-a)\;, & \mbox{step} \\[4mm]
  -i\delta(x-a)\hp\delta(y-a)\;, & p=1 \;. \end{array}\right.
\eeql{sing}
However, for $\mu>0$ I find $\sum_j\omega_jf_j^2(a)e^{-i\omega_j0^+}=-2i(n^-+1)/\mu^2$ (the effect of the regulator can be found using a WKB calculation of the ${\cal O}(j^{-1})$ tail of the summand). Apparently, the highly singular regulated sum (\ref{sing}) with its worse convergence no longer allows commuting the limit $n^-\rightarrow\infty$ with the sum over $j$.

\section{Diagonal two-variable QNM expansion}
\label{diagApp}

Let us try to have a more general look at ``diagonal" expansions such as (\ref{S-chi}) and (\ref{act-phi}), which also occur several times in Ref.\cite{sq}. Thus, suppose one has the {\em one-component\/} expansion
\beq
  \phi(x,y)=\sum_ja_jf_j(x)f_j(y)
\eeql{diag-exp}
for a certain symmetric $\phi$ on $(x,y)\in{[0,a]}^2$. One sees at once that $[\rho^{-1}(x)\partial_x^2-\rho^{-1}(y)\partial_y^2]\hp\phi(x,y)=0$, so that $\phi$ is completely determined by a set of boundary conditions which specify a unique solution to a hyperbolic equation. Hence, $\{f_j(x)f_j(y)\}$ is grossly undercomplete in the space of functions on $0\le x\le y\le a$.

The existence question for the expansion (\ref{diag-exp}) thus has a negative answer in general. However, one has uniqueness under the assumption that $\sum_ja_j$ converges absolutely. For a proof, suppose that $\phi=0$, i.e.,
\beq
  \sum_ja_jf_j(x)f_j(y)=0\;.
\eeql{I}
Operating on (\ref{I}) with $\int_0^{a^+}\!dx\,\rho(x)f_k(x)$ yields
\beq
  \sum_ja_jf_j(y)\left[\delta_{jk}
  -i\frac{f_j(a)f_k(a)}{\omega_j+\omega_k}\right]=0\;.
\eeql{II}
Evaluating (\ref{II}) at $y=a$ gives
\beq
  \sum_ja_jf_j(a)\left[\delta_{jk}
  -i\frac{f_j(a)f_k(a)}{\omega_j+\omega_k}\right]=0\;,
\eeql{III}
while operating on (\ref{II}) with $-i\partial_y|_{a^-}$ shows that
\beq
  \sum_ja_j(\omega_j-i\mu\omega_j^2)f_j(a)\left[\delta_{jk}
  -i\frac{f_j(a)f_k(a)}{\omega_j+\omega_k}\right]=0\;.
\eeql{IV}
Furthermore, $-\rho^{-1}(a^-)\partial_y^2\mbox{(\ref{II})}|_{a^-}$ leads to
\beq
  \sum_ja_j\omega_j^2f_j(a)\left[\delta_{jk}
  -i\frac{f_j(a)f_k(a)}{\omega_j+\omega_k}\right]=0\;,
\eeql{V}
and evaluating (\ref{I}) at $x=y=a$ one gets
\beq
  \sum_ja_jf_j^2(a)=0\;.
\eeql{VI}
Finally, $\omega_k\mbox{(\ref{III})}+\mbox{(\ref{IV})}+i\mu\mbox{(\ref{V})}+if_k(a)\mbox{(\ref{VI})}$ reads
\beq
  2\omega_kf_k(a)\hp a_k=0\quad\Rightarrow\quad a_k=0\;.\qquad\Box
\eeq
The technical condition on the summability of $a_j$ enables the limit $y\uparrow a$ to be taken behind $\sum_j$ to arrive at (\ref{III})--(\ref{V}). Namely, for $\mu>0$ the factor in square brackets is ${\cal O}(j^{-2})$, and this is multiplied at most with $a_j\omega_j^2f_j(y)$ where $f_j(y)$ is bounded. Hence, $\sum_j$ converges uniformly with respect to $y$. If $\mu=0$ the factor in square brackets is ${\cal O}(j^{-1})$, but now the prefactor is at most $a_j\omega_jf_j(y)$ since (\ref{V}) is not needed.

We have not exhaustively examined the intricacies of poorly converging or distributional expansions (\ref{diag-exp}). However, the above theorem and our experience suggest very strongly that the only freedom in the expansion consists in the addition of a term $a_j=c/\omega_j$ if $\mu>0$; since in the course of the derivation we also supposed convergence to $\phi(x,y)$ {\em at\/} the point $x=y=a$, by (\ref{sumtable}) even this freedom is absent if $\mu=0$ (step discontinuity).

 

\begin{references}

\bibitem{optcav} 
R. Lang, M. O. Scully and W.E. Lamb, Phys.  Rev. A {\bf 7}, 1788 (1973); 
R. Lang and M.~O. Scully, Opt. Comm. {\bf 9}, 331 (1973);
H. Dekker, {\it ibid.} {\bf 10}, 114 (1974); Physica C {\bf 83}, 183 (1976);
J. C. Penaforte and B. Baseia, Phys.  Rev. A {\bf 30}, 1401 (1984); 
J. Gea-Banacloche, N. Lu, L. M. Pedrotti, S. Prasad, M. O. Scully and
K. W\'odkiewicz, {\it ibid.} {\bf 41}, 369 (1990); A. J. Campillo, J.
D. Eversole and H.-B. Lin, Phys. Rev. Lett. {\bf 67}, 437 (1991).

\bibitem{meystre} P. Meystre and M. Sargent III, {\it Elements of Quantum Optics\/} (Springer-Verlag, Berlin, 1991).

\bibitem{S&Z} G. Sch\"on and A.D. Zaikin, Phys. Rep. {\bf198}, 237 (1990).

\bibitem{weiss-boek} U. Weiss, {\it Quantum Dissipative Systems} (World Scientific, Singapore, 1993).

\bibitem{grav} 
See, e.g., S. Chandrasekhar, {\it The Mathematical Theory of Black Holes\/} 
(Univ. of Chicago Press, 1991).

\bibitem{comp2}
E. S. C. Ching, P. T. Leung, W. M. Suen, and K. Young,
Phys. Rev. Lett. {\bf 74}, 4588 (1995);
Phys. Rev. D {\bf 54}, 3778 (1996).

\bibitem{HD-rep} H. Dekker, Phys. Rep. {\bf80}, 1 (1981).

\bibitem{KL}P. T. Leung, A. Maassen van den Brink and K. Young, in 
{\it Frontiers in Quantum Physics}, proceedings of the International Conference, Kuala Lumpur, edited by S. C. Lim, R.~Abd-Shukor and K. H. Kwek (Springer-Verlag, Singapore, 1998), p. 214.

\bibitem{sq}K.~C. Ho, P.~T. Leung, A. Maassen van den Brink, and K. Young, Phys. Rev. E {\bf 58}, 2965 (1998).

\bibitem{RMP}E. S. C. Ching, P. T. Leung, A. Maassen van den Brink, W.
M. Suen, S. S. Tong, and K. Young, Rev. Mod. Phys. {\bf 70}, 1545 (1998).

\bibitem{plank}R. W. F. van der Plank and L. G. Suttorp, Phys. Rev. A {\bf53}, 1791 (1996).

\bibitem{diss}
R. P. Feynman and F. L. Vernon, Ann. Phys. {\bf 24}, 118 (1963);
P. Ullersma, Physica {\bf 32}, 27 (1966);
A. O. Caldeira and A.~J. Leggett, Ann. Phys. (N.Y.) {\bf 149}, 374 (1983);
H.~Grabert, U. Weiss and P. Talkner, Z.~Phys. B. {\bf 55}, 87 (1984);
P. S. Riseborough, P.~H\"anggi and U. Weiss, Phys. Rev. A {\bf 31}, 471 (1985).

\bibitem{phi-cont}
Since the positivity of $\rho$
limits its singularity at $x=a$ to at most a $\delta$-function,
$\phi$ itself is continuous. Stronger singularities in $\rho$ would
also leave equations such as (\ref{def-QNM}) undefined
distributionally.

\bibitem{comp1}
P. T. Leung, S. Y. Liu, and K. Young,
Phys. Rev. A {\bf 49}, 3057 (1994); {\bf 49}, 3982 (1994);
P. T. Leung, S.~Y. Liu, S. S. Tong, and K. Young,
{\it ibid.} {\bf 49}, 3068 (1994).

\bibitem{tong}
P. T. Leung, S. S. Tong, and K. Young, J. Phys. A {\bf 30}, 2139 (1997);
{\bf 30}, 2153 (1997).

\bibitem{factor}A. Maassen van den Brink and K. Young, preprint.

\bibitem{bior} P. T. Leung, W. M. Suen, C. P. Sun, and K. Young,
Phys. Rev. E {\bf 57}, 6101 (1998).

\bibitem{jordan}A. Maassen van den Brink and K. Young, preprint math-ph/9905019, submitted to Phys. Rev. E.

\bibitem{Imsource} Since the Euclidean phase-space action does not contain a term $\propto\phi^2$ in the absence of mass regularization, there is no Gaussian $\phi$-cutoff and the source term should be purely imaginary if the path integral is to have a fighting chance of being definable. Due to the overall factor $i$ in the bilinear map, the choice in (\ref{pathint}) satisfies this condition for real $\vec{\chi}$.

\bibitem{schulman} L. S. Schulman, {\it Techniques and applications of path integration\/} (Wiley, New York, 1981).

\bibitem{roepstorff} G. Roepstorff, {\it Path integral approach to quantum physics\/} (Springer-Verlag, Berlin, 1994).

\bibitem{inter} When extending the present work to include interactions as contemplated in Section~\ref{discuss}, the use of phase-space integration should not lead to problems at least in the perturbative regime, since in that case the interacting path integral merely is a formal tool for arriving at the diagram expansion.

\bibitem{HDstring} E.g., H. Dekker, Phys. Lett. {\bf104A}, 72 (1984).

\bibitem{out-exp}The QNM expansion is thus applied to all two-component fields, i.e., not only to outgoing ones, since the path integral is not restricted to the latter, cf.\ the remarks below (\ref{asympt}). For the---easy---justification, see Refs.\cite{sq} (where one faces the analogous issue for operator fields) and\cite{factor}.

\bibitem{conslim}However, this can not be used to incorrectly conclude that the effective action would vanish in the conservative limit. Rather, if $j=-k$ the denominator $\omega_j+\omega_k$ in the action of (\ref{S-QNM}) also tends to zero, and in the limit the diagonal contribution of these terms only yields the action of the closed cavity in terms of its NMs. Cf.\ Ref.\cite{sq}, the end of Section~VII and Appendix~A.

\bibitem{domain}In (\ref{complsq-a}), there seems to be a hidden complication in that $\bar{a}_{jm}^\phs\neq\bar{a}_{-j,-m}^*$ in general, making the integration in $\bar{a}$-space $b$-dependent. However, one can convince oneself (most systematically by splitting the $\bar{a}$-integrals into real and imaginary parts and subsequently using contour methods) that this does {\em not\/} affect the outcome of the path integral, i.e., that the full $b$-dependence in (\ref{complsq-a}) is the one indicated explicitly.

\bibitem{diff-b}Using the standard shorthand $\partial_b\equiv\frac{1}{2}[\partial_{\re b}-i\partial_{\im b}]$ (and $\partial_{b^*}\equiv\frac{1}{2}[\partial_{\re b}+i\partial_{\im b}]$), one can manipulate the $b_{jm}$---obeying $b_{jm}^\phs=b_{-j,-m}^*$---formally as if they were independent variables.

\bibitem{ryder} L. H. Ryder, {\it Quantum Field Theory\/} (Cambridge University Press, Cambridge, 1986).

\bibitem{AGD} 
A. A. Abrikosov, L. P. Gor'kov and I. E. Dzyaloshinski, 
{\it Methods of Quantum Field Theory in Statistical Physics\/} 
(Dover, New York, 1975).

\bibitem{partII}
K. C. Ho, P. T. Leung, A. Maassen van den Brink and K. Young, in {\it Proceedings of the APPC7 conference}, edited by Hesheng Chen (Science Press, Beijing, 1999), p. 433;
P.~T. Leung, A.~Maassen van den Brink and K. Young, in preparation (1999).

\bibitem{tensorF}
We wish to take the opportunity to amend an unnecessarily complicated argumentation in Ref.\cite{sq}. Namely, defining the tensor correlator $\tilde{\sf F}(x,y,\omega)\equiv\langle\tilde{\vec{\phi}}(x,\omega)\otimes\vec{\phi}(y)\rangle$ as in (6.6) (all references in this note are to Ref.\cite{sq}), a key point of \cite{sq} is that the field operators $\vec{\phi}$ have QNM expansions (2.8) just like their classical counterparts. Substitution of these into (6.6) leads to $\tilde{\sf F}(x,y,\omega)=\sum_{jk}\langle\tilde{a}_j(\omega)\hp a_k\rangle\fj(x)\otimes\vec{f}_{\hn\!k}(y)$, and (6.2), (6.3) now at once yield (6.8) for $\tilde{a}_{jk}$ upon comparison with (6.7). While this simplified calculation does not use the material of Appendix~C any more, the tensor expansion presented in the latter remains correct, and is useful for reference.

\bibitem{Gsign} However, the Green's functions $G^{\rm R}$ of this paper and $G$ of Ref.\cite{jordan} have opposite signs, since the customary sign for the retarded quantum propagator is not the most convenient one in classical field theory. The signs of the associated Wronskians differ accordingly.


\end{references}
\end{document}